\documentclass[twocolumn,aps,showpacs,amsmath,amssymb]{revtex4-2}
\usepackage{amsmath,amssymb}
\usepackage{citesort}
\usepackage{graphicx}
\begin{document}

\title{Phase separation of passive particles in active liquids}
\author{Pragya Kushwaha$^{1}$, Vivek Semwal$^2$, Sayan Maity$^1$, Shradha Mishra$^{2}$, Vijayakumar Chikkadi$^{1}$}

\affiliation{
$^{1}$ Indian Institute of Science Education and Research Pune, India 411008\\
$^{2}$ Indian Institute of Technology (BHU) Varanasi, India 221005\\
}

\date{\today}

\begin{abstract}
The transport properties of colloidal particles in active liquids have been studied extensively. It has led to a deeper understanding of the interactions between passive and active particles. However, the phase behavior of colloidal particles in active media has received little attention. Here, we present a combined experimental and numerical investigation of passive colloids dispersed in suspensions of active particles. Our study reveals dynamic clustering of colloids in active media due to an interplay of active noise and an attractive effective potential between the colloids. The size-ratio of colloidal particles to the bacteria sets the strength of the interaction. As the relative size of the colloids increases, the effective potential becomes stronger and the average size of the clusters grows. The simulations reveal a macroscopic phase separation of passive colloids at sufficiently large size-ratios. We will present the role of density fluctuations and hydrodynamic interactions in the emergence of effective interactions.
\end{abstract}


\maketitle

The Brownian colloids self-assemble to display a wide variety of phases depending on their shapes and interactions \cite{Sacanna13, Sacanna21, Glotzer07}. Their equilibrium phase behavior is governed by the principles of equilibrium statistical mechanics \cite{Pusey91, Poon15}. However, our understanding of the collective behavior of colloids far from equilibrium remains a challenge \cite{Gokhale22,Kummel15}. In recent years, active matter has emerged as a new paradigm for understanding nonequilibrium systems \cite{Romanczuk12, Marchetti13, Ramaswamy17,Bechinger16}. They are known to display many interesting phenomena such as flocking \cite{Vicsek95, Bricard13}, motility induced phase separation \cite{Cates15-1,Buttinoni13,Palacci2013}, active turbulence \cite{Wensink12}, superfluidity \cite{Lopez15}, that are absent in equilibrium systems. Therefore, active matter offers novel approaches to colloidal assembly in systems far from equilibrium. In this letter, we have investigated the phase behavior of colloidal particles dispersed in active liquids. 

Wu and Libchaber \cite{Wu00} did seminal experiments on the active transport of colloidal particles in suspensions of bacteria. They discovered anomalous diffusion and a large effective diffusion constant, when compared to diffusion at equilibrium, which inspired a slew of theoretical investigations and detailed experiments \cite{Leptos09, Valeriani11, Tiffeault15, Polin16, Polin18, Yeomans17, Ortlieb19, Gibaud20}. The subsequent efforts have elucidated how enhanced diffusion arises due to an interplay of entrainment of colloids by bacteria, far-field hydrodynamic interactions, direct collisions, and the relative size of bacteria and colloid \cite{Polin16, Polin18, Yeomans17}. Further, the effective interaction between a pair of passive particles in active media has been the focus of several investigations. It has been predicted to be attractive, repulsive, and long-ranged, depending on the geometry of passive particles, the activity of active species, and their density \cite{Leonardo11, Reichhardt14, Ni15, Ali-Naji17, Zhao21, Lowen15, Cacciuto14, Kafri18, Yang20}. This understanding has opened new routes to colloidal assembly mediated by active fluids \cite{Kummel15, Brady19}. The phase behavior of active-passive mixtures is a topic of recent interest \cite{Hagan12, Brady15, Cates15-2, Frey16, Gompper16, Dolai18, Brady19, Joanny20, Liebchen20, Bechinger16}, where experimental investigations are scarce \cite{Kummel15, Gokhale22}. On the one hand, theory and simulations at high Peclet numbers have shown that homogeneous mixtures of active and passive particles are unstable. The underlying physics is similar to motility induced phase separation (MIPS) \cite{Cates15-2,Gompper16}. On the other hand, in the diffusive limit, theory and simulations of nonequilibrium binary mixtures with different diffusivities and temperatures reveal phase separation \cite{Joanny15,Frey16,Joanny20} due to spinodal-like instability. Surprisingly, there is little known about mixtures at moderate Peclet numbers. This is the range where most of the active matter experiments involving living matter or synthetic systems, such as diffusio-phoretic colloids, fall. A recent study of colloids in active suspensions of bacteria reports dynamical clustering and absence of phase separation at moderate Peclet numbers \cite{Gokhale22}. The conclusions were based on the phase diagram obtained from variations of Peclet number and rotation rate of active particles. In contrast, earlier numerical studies have shown a macroscopic phase separation \cite{Dolai18}. Therefore, it is not clear whether active-passive mixtures show macroscopic phase separation at moderate Peclet numbers. 

This letter presents a combined experimental and numerical study of the phase behavior of colloidal particles in active liquids. The experiments were performed using colloids in bacteria suspensions, and simulations of active-passive mixtures were realized using Brownian dynamics \cite{Schilling,Schuss,Fily12}. Earlier simulations of active-passive mixtures, by one of the authors of this letter, had shown a significant influence of the size-ratio of passive to active particles on their phase diagram \cite{Dolai18}. Motivated by this study, our experiments were performed over a range of densities and sizes of passive colloidal particles in active suspensions of bacteria. The colloids display dynamic clustering due to an interplay of activity and an attractive effective potential. However, the average size of the clusters increases with the size of colloidal particles, suggesting an enhanced interaction between the particles. Using simulations, we confirm an attractive effective potential between passive particles in an active medium. The strength of the interaction is shown to grow with an increasing size-ratio. When the size-ratio is sufficiently large, the interactions are strong enough to drive the phase separation of passive colloids. The origin of the effective potential in simulations appears to be related to long-ranged density fluctuations of active particles. In contrast, the correlations of density fluctuations of bacteria decay rapidly in experiments. These results indicate a hydrodynamic origin of effective interactions between the colloids in our experiments. Thus, shedding new light on the phase behavior of passive particles in active media. 

The active suspensions were prepared using E.coli cells (U5/41 type strain). The cells were cultured using well established protocols in the literature \cite{Adler67,Lopez15}. They are suspended in a motility media to get desired concentrations. Details of the method are given in the supplementary section. The density of bacteria in our experiments was well below the density threshold for the onset of collective motion. The average speed and average size of bacteria cells were estimated to be $v=33.84\pm9.98~\mu m/s$ [supplementary Fig.~S1(a)] and $l=2.68\pm0.86~\mu m$ [supplementary Fig.~S1(b)], respectively. Their rotational diffusion time scale was estimated to be $\tau_r=1.67~s$ [supplementary Fig.~S1(c)]. The Peclet number, which is defined as $Pe=\tau_r v/l$, turns out to be $Pe\sim 21$ for our system. The phase behavior of colloidal particles in suspensions of bacteria was investigated by varying the size and density of the beads at a constant density of bacteria. The diameters of the particles used in the experiments were $\sigma= 7\mu m,~ 10\mu m ~\textnormal{and} ~15\mu m $, and their density is varied from $\phi\sim 0.1-0.4$, where $\phi=N*\pi\sigma^2/(4A)$ is the area fraction, $N$ is the number of colloidal particles in the field of view of area $A$. The size-ratio $S=\sigma/l$ is defined as the ratio of the diameter of colloids to the length of the bacteria. 

The simulations were performed using a binary mixture of active Brownian particles (ABP) with $N_a$ small active particles of radius $a_a$ and $N_p$ big passive particles of radius $a_p$ ($a_p > a_a$) moving on a two dimensional frictional substrate. The active particles are associated with a self propulsion speed ${v}$ and an orientation unit vector $\hat{v}_i$. The equations of motion and other simulation details are given in the supplementary section. We simulate the system in a square box of size $l_{box} \times l_{box}$, with periodic boundary conditions. The system is defined by the area fractions $\phi_a= N_a\pi a_{a}^2/l_{box}^{2}$ and $\phi_p= N_p \pi a_{p}^{2}/l_{box}^{2} $ of the active and passive particles respectively, the activity $v$ of active particles and the size-ratio $(S=a_p/a_a)$ defined as the ratio of the radius of a passive particle to the radius of an active particle. 
We start with a random homogeneous distribution of active and passive particles in the box and with random directions for the velocity of active particles. The Eqs.~(S1-S3) are updated for all particles and one simulation step is counted after a single update for all the particles. The simulations does not include the hydrodynamic interactions that are present in experiments. The effect of hydrodynamic interaction can be included using coarse-grained studies similar to \cite{Catesprl2015}.

\begin{figure}[h]
\includegraphics[width=0.4\textwidth]{ 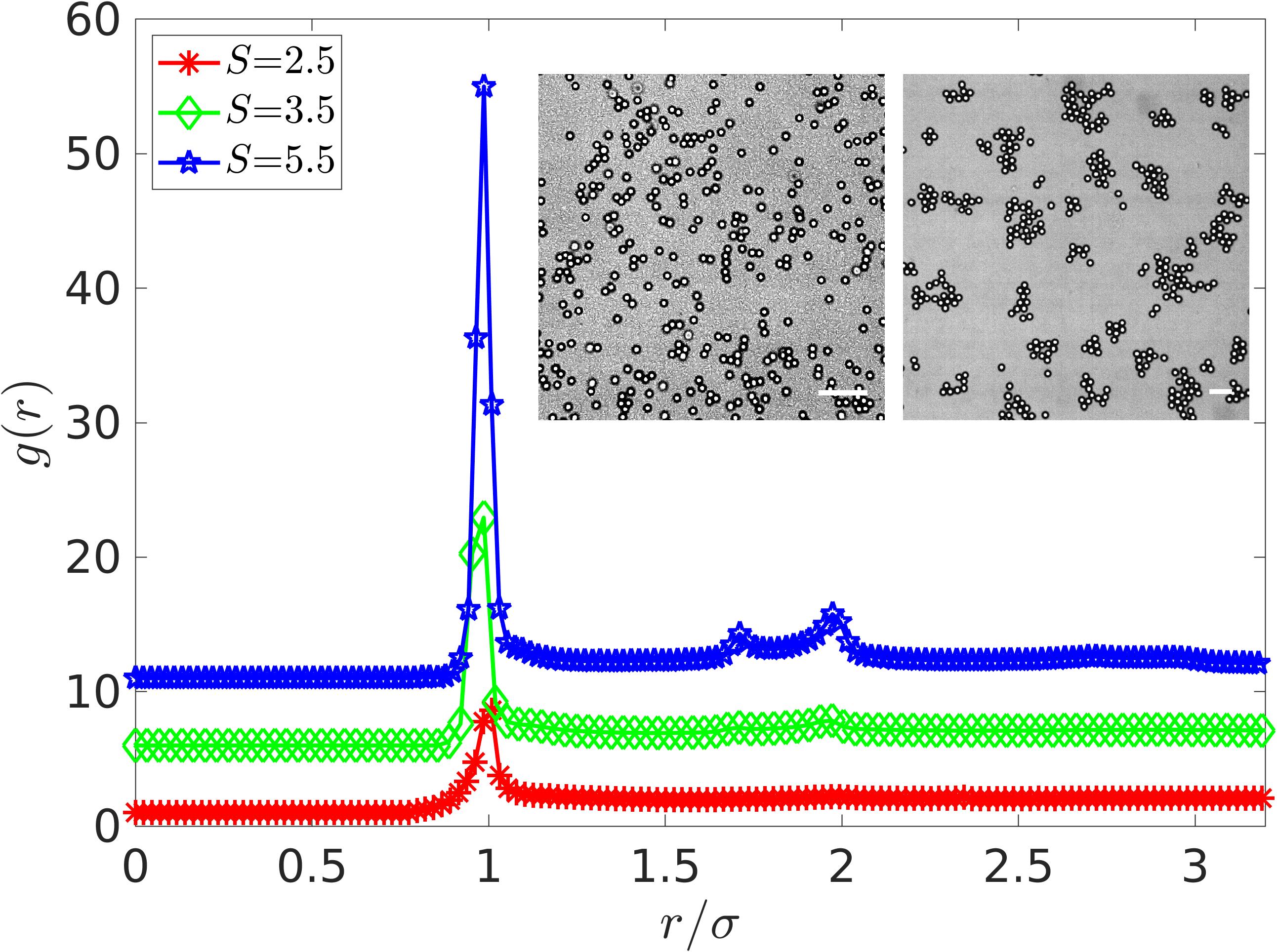} 
\caption{The structure of passive particles in active suspensions. Main panel : The pair correlation function $g(r)$ for $\phi\sim 0.10$ and $S\sim2.5,~3.5, ~\text{and}~5.5$. The $g(r)$ curves are shifted along the $y-axis$ for clarity. Insets: The bright field images of particles at $\phi\sim0.10$ and size ratios $S\sim2.5$ (left) and $S\sim5.5$ (right), respectively. The scale bar in the images is $50\mu m$. }
\label{fig1}
\end{figure}

The colloids used in our experiments are bigger than $5~\mu m$, so they are non-Brownian particles. However, they diffuse in suspensions of bacteria due to active fluctuations with a characteristic super-diffusive motion on short time scales and a diffusive motion on long time scales. To investigate their collective behavior in active suspensions, we first analyze their pair correlation function $g(r)$, which is shown in the main panel of Fig.~1 at an area fraction of $\phi\sim 0.1$ and size ratios $S\sim2.5-5.5$. The normalized $g(r)$ for different size-ratios is shifted along the y-axis for clarity. What is prominent is the presence of a sharp peak at $r=\sigma$, and additional peaks develop at $r=1.7\sigma$ and $r=2\sigma$ with increasing size ratio. The peak at $2\sigma$ indicates a second shell of neighbors, and the one at $1.7\sigma$ is a signature of hexagonal ordering in the cluster. These observations are evident in the bright field images presented in the insets of Fig.1. The larger size ratios lead to larger clusters with enhanced order. These images are reminiscent of clustering in systems of purely active particles \cite{Buttinoni13}. However, the clusters of passive particles in our experiments break and form much more rapidly. A real-time video of dynamic cluster formation is presented in the supplementary section Video.~SV1 for $\phi\sim0.10$ and $S\sim2.5$. Recent simulations have reported similar dynamic clustering and traveling interfaces of active-passive particles that are not observed in our current study \cite{Cates15-2, Gompper16}. One of the main difference between our experiments and these simulations are the large Peclet numbers used in simulations. Further, as reported by earlier investigations, self-propulsion of particles is a manifestation of an attractive effective potential between the passive particles due to active fluctuations \cite{Leonardo11}. 

\begin{figure}[!]
\includegraphics[width=0.23\textwidth]{ 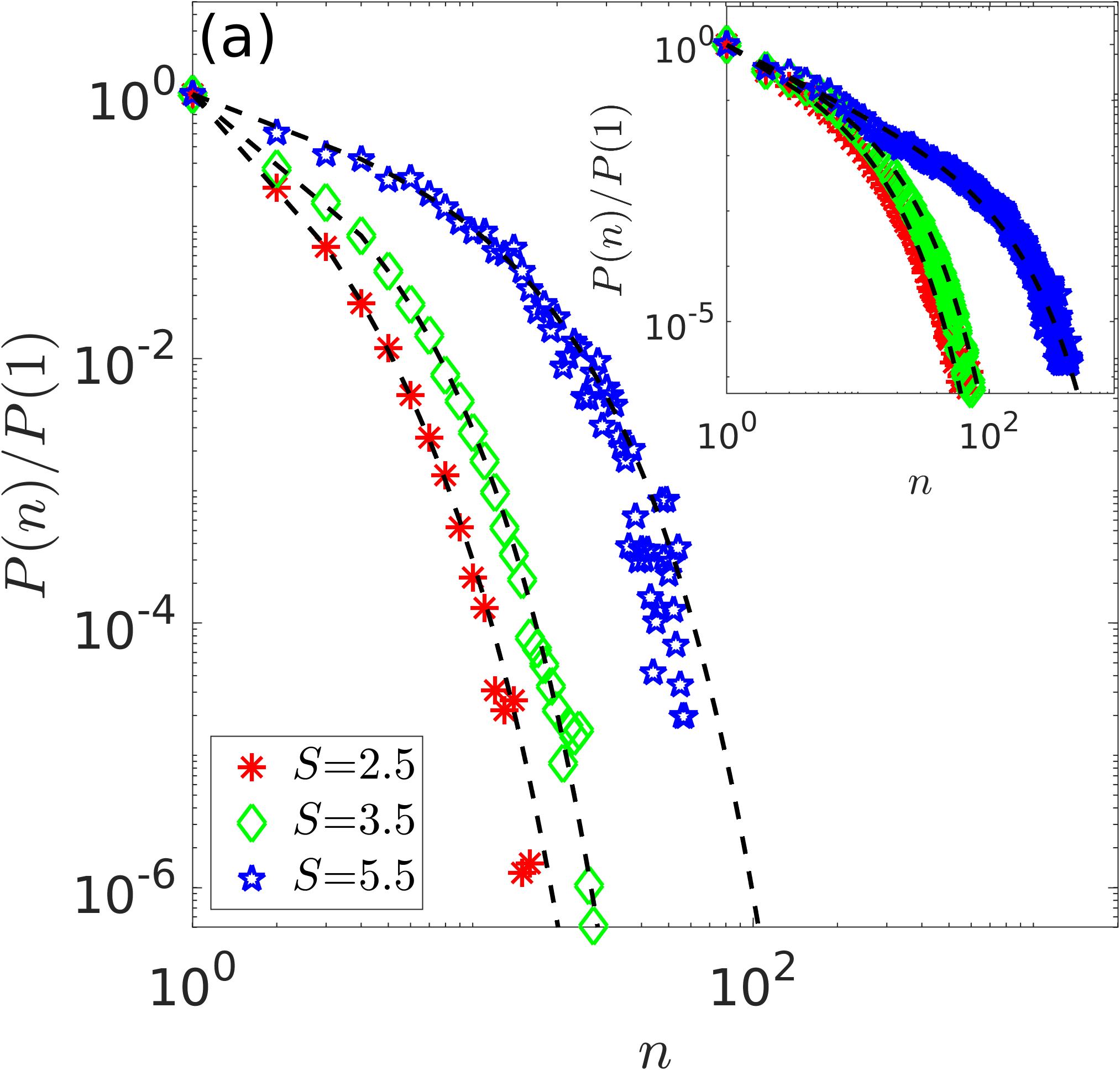} 
\includegraphics[width=0.225\textwidth]{ 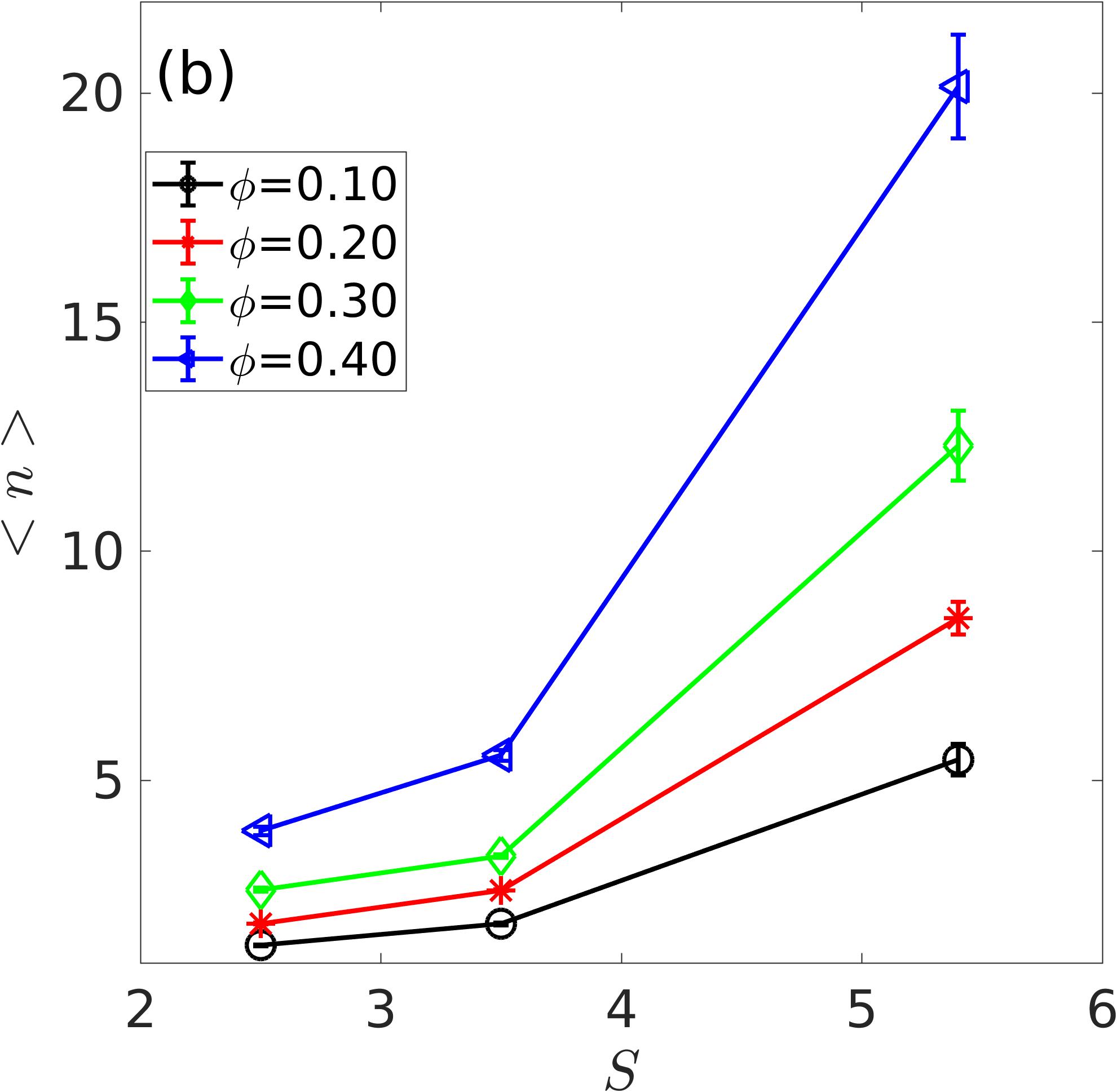}
\caption{Cluster statistics of passive particles. (a) Cluster size distribution in the main panel is shown for different size-ratios $S\sim2.5,~3.5,~\text{and}~5.5$ at a density of $\phi\sim0.10$. The symbols distinguish different size ratios. The inset shows the CSD plot for same size ratios at $\phi\sim0.3$. (b) The average cluster size $<n>$ for varying $S$. The curves with different symbols correspond to different particle densities, ranging from $\phi\sim0.1-0.4$.}
\label{fig2}
\end{figure}
We next turn our attention to cluster size distribution (CSD), $p(n)$, which is a count of clusters of $n$ particles \cite{Dolai18,csd}. The clusters in our experiments were determined by setting a distance criterion of $r_c\leq 1.1\sigma$ to identify pairs of particles as neighbors. This was set based on the position of the first peak of $g(r)$ in Fig.1, and to account for small polydispersity ($<5\%$) in the size distribution of colloidal particles. The results of our analysis are presented in Figs.~2a \& 2b. The main panel in Fig.2a shows CSD for varying size ratios of $S\sim2.5,~3.5,~\text{and}~5.5$ at a density of $\phi\sim0.1$. For small size ratios $S<5$, $p(n)$ has an exponential form $exp(-n/n_0)$ as observed in the equilibrium case \cite{Fily12}. The clustering is weak at these size ratios, however, for $S>5$ the $p(n)$ displays a power-law decay with an exponential cut-off at large $n$, i.e., it is best described by $p(n)/p(1)\sim 1/{n}^{\alpha}~exp(-n/n_0)$. The fits of this form to our data are shown in the figure using dashed lines. These results indicate that the characteristic size of clusters grows with increasing size ratio. The growth of clusters is dramatic at larger area fractions, the inset of Fig.2a shows cluster distribution at $\phi\sim0.3$. 

We further elucidate the clustering of colloids by computing the average cluster size using the expression $<n>=\sum~n~p(n)$, which is presented in Fig.~2b where the curves with different symbols correspond to different area fractions ranging from $\phi\sim 0.1-0.4$. These measurements were made in the steady state where the mean cluster sizes fluctuates around a mean value. This data is provided in Fig.~S2(a-c) for various size ratios and area fractions for over $5000$ frames or $500~s$. What is clear from Fig.~2b is that increasing the size-ratio or the relative size of colloids leads to larger cluster sizes. This suggests that the effective potential between the colloids becomes stronger with an increasing size ratio. One can intuitively understand the underlying physics by considering the interaction between an isolated colloidal particle and a swimmer. When the size of a particle is small, a bacterium entrains the particle to larger distances before changing its direction of motion. However, when the particle is large, the entrainment distance is small, and the scattering angle of the swimmer is large \cite{Yeomans17}. It indicates that the bacteria can suppress cluster formation when the colloidal particles are smaller. What is not clear from our experiments is whether larger size-ratios lead to a macroscopic phase separation in our system. To understand this aspect, we turn to numerical simulations that allow a detailed exploration of parameter space. 

The first quantity we have calculated in the simulations is the effective potential between two passive particles in the medium of ABPs with torque. In order to calculate the effective potential between two passive particles, we choose $N_p=2$ at positions ${\bf r}_1$ and ${\bf r}_2$, respectively, in a system of ABPs with $N_a=1800$. We keep ${\bf r}_1$  fixed and slowly vary ${\bf r}_2$ in small steps of $\Delta x = 0.5 a_a$ starting from the zero surface to surface distance between two passive particles. The cartoon of the system simulated for the force calculation for a fixed $r$ is shown in Fig.~S3 ({\bf SM}). In the figure, ABPs are shown in red and passive particles in blue for $S=8$. For resolution, only a part of the system near the two passive particles is shown. The active particles' positions and orientations are updated according to the Eqns.~(S1 and S2). For each configuration at a given distance between two passive particles, the system is allowed to reach the steady state. Further, we use the steady state configuration to calculate the force ${\mathcal{ F}}^{S}(r)$ between two-passive particles at a surface to surface separation $r$, such that ${\mathcal{F}}^{S}(r)={\bf{F}}_{12}(r)+\sum_{i=1}^{N_a}{\bf{F}}_{1i}(r)$. 
Here ${\bf{F}}_{12}(r)$ is the force due to passive particle $2^{nd}$ on $1^{st}$, and $\sum_{i=1}^{N_a}{\bf{F}}_{1i}(r)$ 
represents the sum of all the forces due to active particles on $1^{st}$ passive particle for a given configuration of two 
passive particles at separation $r$. The potential is then calculated by integrating the force over the distance 
$U(r) = \int_{-\infty}^{r}{\bf \mathcal{F}}^{S}(r)dr$  \cite{jaydeb,jaydeb1,jaydeb2}. Here we set the lower limit as one-fourth of the box-length. The results are averaged over $30$ independent realizations.

We calculated the effective potentials $U(r)$ for $Pe= 25$ (which is comparable to  the experimental system) and four size-ratios $S=3$, $5$, $8$ and $10$. The comparable size-ratio in experimental system is $S \sim (2.5$ to $5.5)$. We first plot the effective potential $U(r)$. In the main panel of Fig.~3(a) we show the plot of $U(r)$ vs. $r$ for the system for $S=3,$ $5$, $8$ and $10$. The distance is normalised by the radius of active particles, which is kept fixed to $0.1$. The negative side of the potential shows attraction and the positive nature is repulsion. For all the parameters the potential approaches zero at large distances, and it is negative at intermediate distances. The depth of the potential becomes deeper with increasing $S$. The inset shows the effective potential with the distance $r$ scaled by the size of passive particles. Surprisingly, the minima of the potentials for the size ratios $S=5$, $8$ and $10$ fall at $r/a_p = 1$, which implies that the length scale characterizing the range of the interaction potential is set by size passive particles. We investigate further the origin of long-range interactions by considering a single passive particle in the center of our system, as shown in Fig.~S4. It is evident that the passive particle disturbs the density field of active particles, leading to clustering around the passive particle. The main panel of Fig.~3(b) shows the normalised density correlation $C(r)=(\left<\rho(0)\rho(r)\right> - \left<\rho(r)\right>^2) / (\left<\rho(r)^2\right>-\left<\rho(r)\right>^2)$ of active particles calculated from the surface of the big passive particle for four different size ratios $S=3$, $5$, $8$ and $10$. The inset of Fig.~3(b) shows the typical size of clusters $L(S)$ around a single passive particle in the center of the box for different size ratios $S$. The $L(S)$ is measured in terms of size of ABP. Clearly, $L(S)$ increases with increasing $S$. The number fluctuations of active particles around an isolated passive particle yield similar conclusions. The Fig.~S5 (SM) shows number fluctuations for three different sizes of the passive particle or for three different size ratios $S=3$, $5$ and $8$. The details of the calculations are give in the supplementary material. For all the cases the graph is a power law with $\Delta N \simeq N^{\alpha}$, where $\alpha \simeq 0.7$ for moderate $N$ for all $S$ and starts to deviate for large $N$. The deviation appears at relatively larger $N$ on increasing size ratio. Hence increasing the size of passive particle increases the stretch of density fluctuation of ABP's. These results establishes that the density fluctuations play a central role in the emergence of long-range effective attractive interactions between passive particles in our simulations.

\begin{figure}[!]
\centering
\includegraphics[width=0.21\textwidth]{ 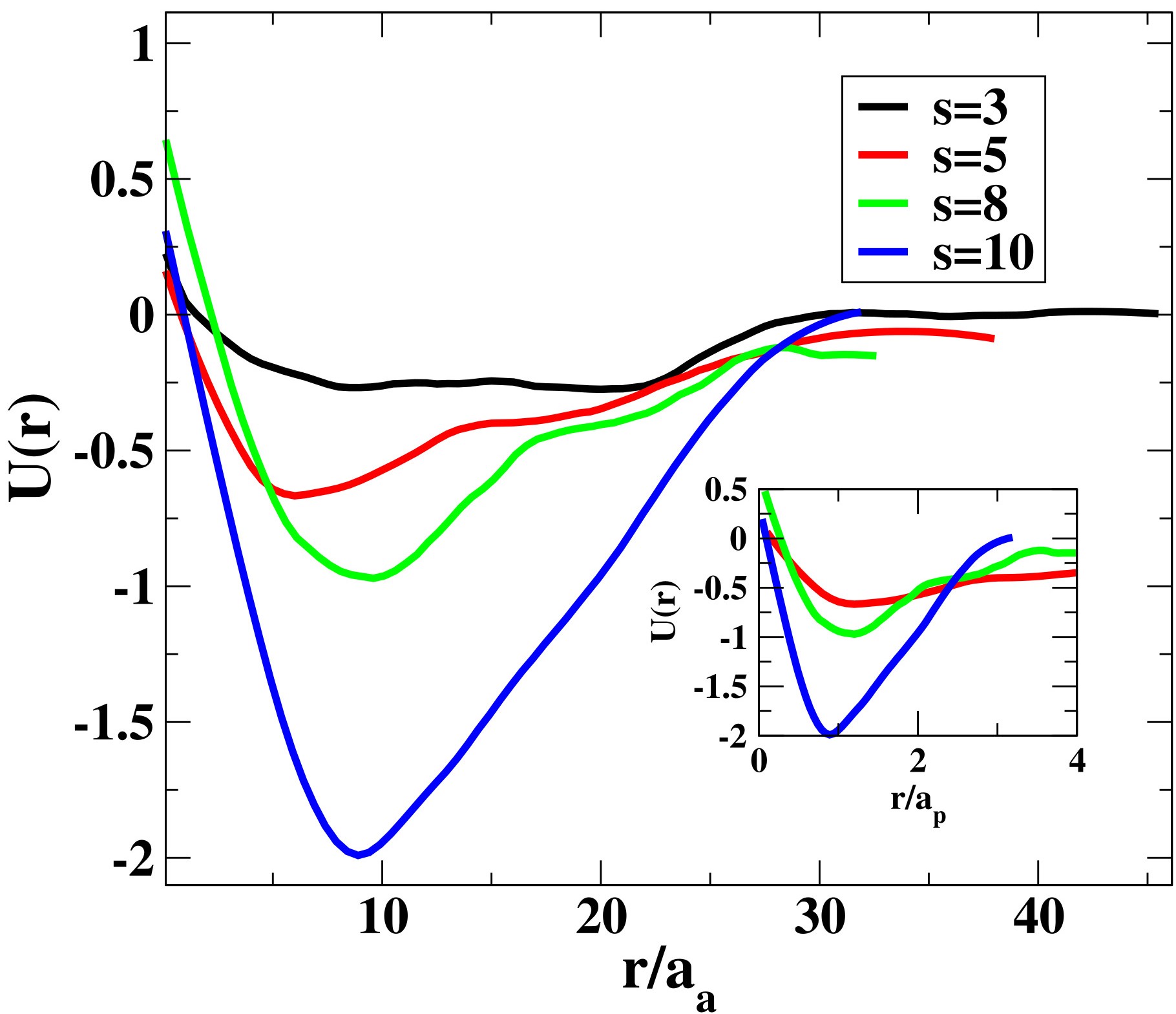}
\includegraphics[width=0.20\textwidth]{ 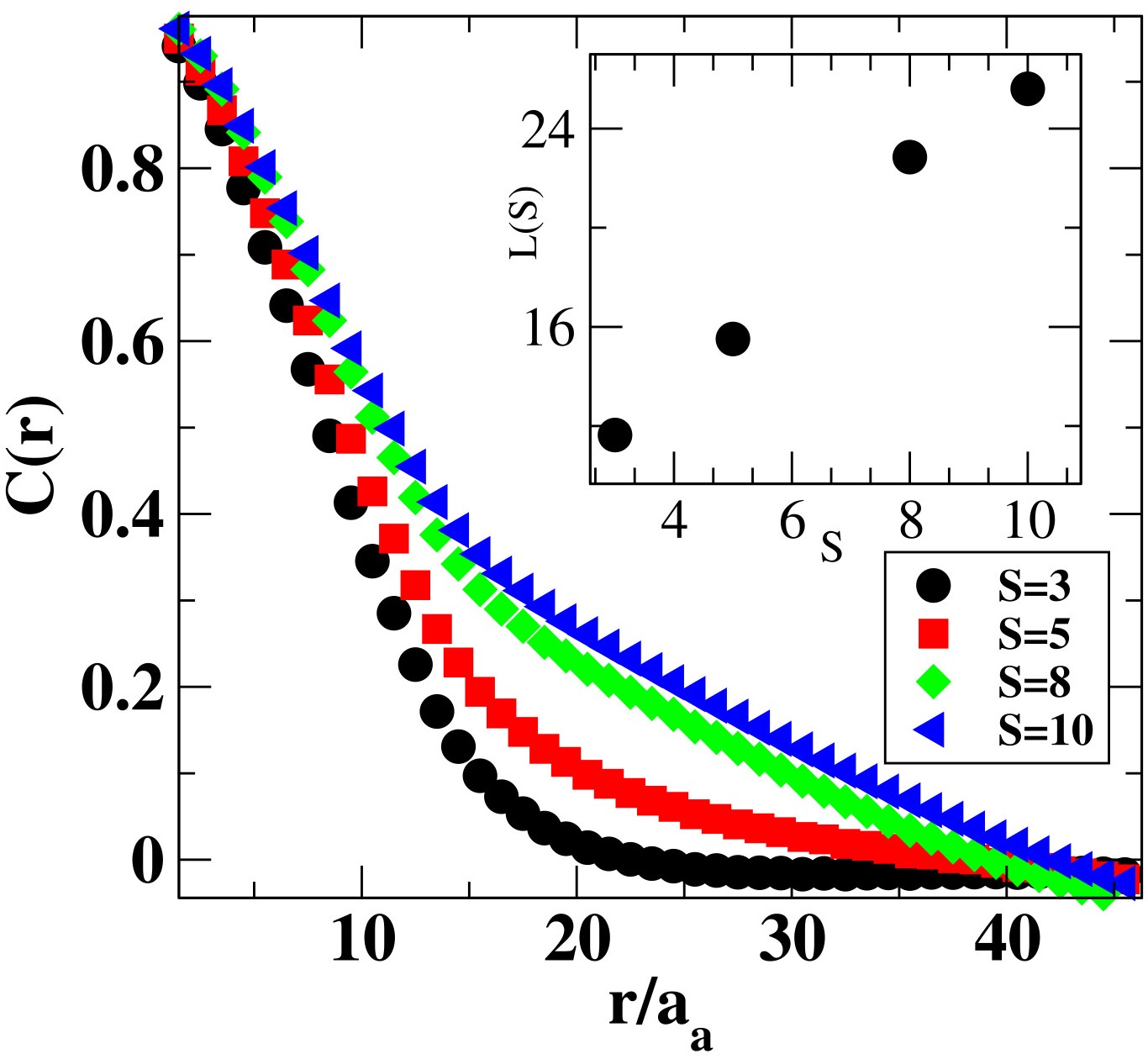}\\
\includegraphics[width=0.4\textwidth]{ 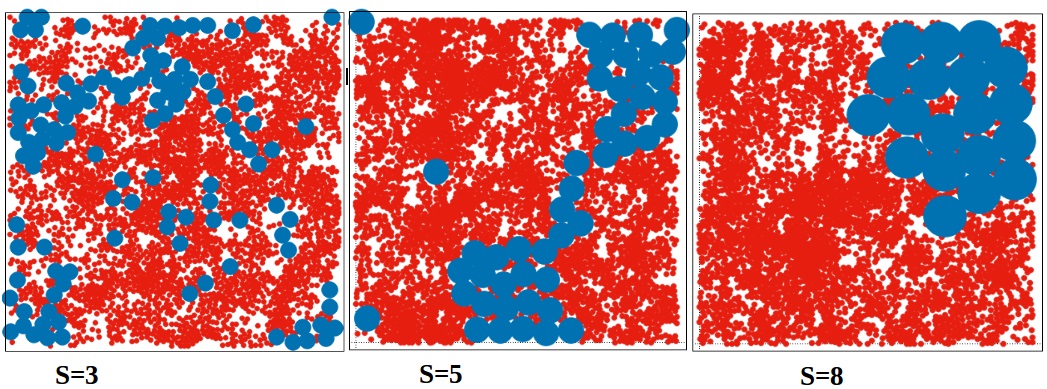}
\caption{Top left figure: The effective potential between a pair ,of colloidal particles. The main panel shows the plot of the effective potential $U(r)$ vs. distance $r$ for $Pe=25$ and size-ratios $S=3,~5,~8, \textnormal{and}~10$. The inset shows the effective potential with the scaled distance $r/a_p$. Top right figure: The main panel shows normalized correlations of density fluctuations $C(r)$ due to a passive particle.  The inset show the length scale extracted from $C(r)$ as function of the size-ratio. The length scale is expressed in terms of active particle size. Bottom panel: Snapshots of the system obtained from the microscopic simulation: two types of particles for different size ratio $S=3,5$ and $8$ (left, central and right columns) at $Pe=25$. Red particles are ABPs and blue particles are passive particles, for fixed packing fraction $\phi=0.60 $  in a system of size $l_{box}=140 a_a$.}
\label{simRes}
\end{figure}

We elucidate the effect of such effective potential, full microscopic simulations of mixtures of active and passive were performed using the Eqs.~(S1-S3). We simulated the system for  $Pe = 25$  and size ratio $S=3, 5$ and $8$, which are close to experimental values. In the bottom panel of Fig.~\ref{simRes} the steady state snapshots of passive (blue, bigger) and active (red, smaller) are shown for different size ratios $S=3$, $5$ and $8$ respectively. Clusters with  moderate to strong ordering is found on increasing $S$. For small  $S=3$ clusters are present but without strong local hexagonal ordering, whereas as we increase $S$ the ordering and clustering is enhanced. We also calculated the percent of passive particles  participating in the largest cluster for different size ratios and it increases from $35\%$ to $67\%$ as we increase size from $3$ to $8$ (data not shown). Hence for large size ratio passive particles show the macroscopic phase separation.

A similar examination of correlations of density fluctuations of bacteria in experiments reveals that they are suppressed, which is evident from $C(r)$ in Fig.~S6. The clustering of colloidal particles appears to arise from their hydrodynamic interactions. An earlier numerical study of active-passive matter with pusher-type swimmers at dilute concentrations had shown hydrodynamic interactions to stabilize colloidal clusters \cite{Garcia15}. In addition, a recent theoretical model of active gels shows a long-ranged attractive effective potential between colloids due to hydrodynamic effects \cite{Madan19}. Considering these studies, hydrodynamics is likely to promote the formation of colloidal clusters.

Our investigations conclude that the interplay of effective potential and active noise determines the phase behavior of colloidal particles in active liquids. The strength of the effective potential is set by the size ratio of passive particles to active ones; larger size ratios lead to stronger interactions. The simulations reveal a long-ranged effective potential extending to several active particle diameters. It appears to emerge from the long-ranged density fluctuations of active particles in the system. When the size-ratio is small, the passive particles display dynamic clusters that form and break rapidly. However, the effective potential is strong enough to lead to phase separation of passive particles at sufficiently large size-ratios. These are the novel features of active-passive mixtures absent in the equilibrium analog of colloid-polymer mixtures where the range of effective potential is short-ranged. The density fluctuations of Bacteria are suppresses in experiments. Further investigation is needed to understand the role of hydrodynamic interactions on the effective potential of colloids in our experiments with active suspensions.

We thank Chaitanya Athale, Apratim Chatterji, Thomas Pucadyil, Sunish Radhakrishnan, Rajesh Singh, and Ganesh Subramanian for helpful discussions and support. We thank Madan Rao for drawing our attention to \cite{Madan19}, and Kumar Gourav for assistance in the initial stages of experiments. V.C. acknowledges financial support from IISER Pune and DST/SERB under the project grant CRG/2021/007824. P.K. is supported by CSIR-UGC fellowship 1353. V.S. and S. M.  thank I.I.T. (BHU) Varanasi computational facility. V.S. thanks DST INSPIRE (INDIA) for the research fellowship. S.M. thanks DST, SERB (INDIA), Project No. ECR/2017/000659 for partial financial support.


\begin{thebibliography}{99}
\bibitem{Sacanna13} S. Sacanna, M. Korpics, K. Rodriguez et al., Shaping colloids for self-assembly, Nat. Commun. 4, 1688 (2013).

\bibitem{Sacanna21} T. Hueckel, G.M. Hocky,  and S. Sacanna, Total synthesis of colloidal matter, Nat Rev Mater 6, 1053–1069 (2021). 

\bibitem{Glotzer07} S. Glotzer and M. Solomon, Anisotropy of building blocks and their assembly into complex structures, Nature Mater. 6, 557–562 (2007).

\bibitem{Pusey91} P. N. Pusey, in Liquids, Freezing and Glass Transition edited by J. P. Hansen, D. Levesque, and J. Zinn-Justin (North-Holland, 1991), pp. 765–942.

\bibitem{Poon15} W. C. K. Poon, “Colloidal suspensions,” in Oxford Handbook of Soft Condensed Matter, edited by E. Terentjev and D. A. Weitz (Oxford University Press, Oxford, 2015), pp. 1–49.

\bibitem{Gokhale22} S. Gokhale, J. Li, A. Solon, J. Gore, and N. Fakhri, Dynamic clustering of passive colloids in dense suspensions of motile bacteria, Phys. Rev. E, 105 (5), 054605 (2022).

\bibitem{Kummel15} F. Kummel, P. Shabestari, C. Lozano, G. Volpe, and C. Bechinger, Formation, compression and surface melting of colloidal clusters by active particles, Soft Matter 11, 6187 (2015).

\bibitem{Romanczuk12} P. Romanczuk, M. Bar, W. Ebeling, and B Lindner, Active Brownian particles , Eur. Phys. J. Special Topics 202,1-162(2012)

\bibitem{Ramaswamy17} S. Ramaswamy, Active matter, Journal of Statistical Mechanics: Theory and Experiment, 2017,054002,(2017)

\bibitem{Marchetti13} M. C. Marchetti, J.-F. Joanny, S. Ramaswamy, T. B. Liverpool, J. Prost, M. Rao, and R. A. Simha, Hydrodynamics of soft active matter, Rev. Mod. Phys. 85, 1143 (2013).

\bibitem{Bechinger16} C. Bechinger, R.D. Leonardo, H. Lowen, C. Reichhardt,
G. Volpe, G. Volpe, Active particles in complex and crowded environments, Rev. Mod. Phys. 88, 045006 (2016).

\bibitem{Vicsek95} T. Vicsek, A. Czirók, E. Ben-Jacob, I. Cohen, and O. Shochet, Novel type of phase transition in a system of self-driven particles, Phys. Rev. Lett. 75, 1226 (1995). 

\bibitem{Bricard13} A. Bricard, J.-B. Caussin, N. Desreumaux, O. Dauchot, and D. Bartolo, Emergence of macroscopic directed motion in populations of motile colloids, Nature 503, 95 (2013).

\bibitem{Cates15-1} M. E. Cates and J. Tailleur, Motility-Induced Phase Separation, Annu. Rev. Condens. Matter Phys. 6, 219 (2015). 

\bibitem{Buttinoni13} I. Buttinoni, J. Bialké, F. Kümmel, H. Löwen, C. Bechinger, T. Speck, Dynamical clustering and phase separation in suspensions of self-propelled colloidal particles, Phys. Rev. Lett. 110, 238301 (2013).

\bibitem{Palacci2013} J. Palacci, S. Sacanna, A. P. Steinberg, D. J. Pine and P. M. Chaikin, Living Crystals of Light-Activated Colloidal Surfers, Science, 339, 936 (2013).

\bibitem{Wensink12} H. Wensink et al., Meso-scale turbulence in living fluids, Proc. Natl. Acad. Sci. USA 109, 14308 (2012).

\bibitem{Lopez15} H.M. Lopez, J. Gachelin, C. Douarche, H. Auradou, and E. Clement, Turning bacteria suspensions into superfluids, Phys. Rev. Lett. 115, 028301 (2015).

\bibitem{Wu00}X.-L. Wu and A. Libchaber, Particle Diffusion in a Quasi-Two-Dimensional Bacterial Bath, Phys. Rev. Lett. 84, 3017 (2000).

\bibitem{Leptos09} K.C. Leptos, J.S. Guasto, J.P. Gollub, A.I. Pesci, and R.E. Goldstein, Dynamics of enhanced tracer diffusion in suspensions of swimming eukaryotic microorganisms,
Phys. Rev. Lett. 103, 198103 (2009).

\bibitem{Valeriani11} C. Valeriani, M. Li, J. Novosel, J. Arlt, and D. Marenduzzo, Colloids in a bacterial bath: simulations and experiments, Soft Matter 7, 5228 (2011). 

\bibitem{Tiffeault15} J.-L. Thiffeault, Distribution of particle displacements due to swimming microorganisms, Phys. Rev. E 92, 023023 (2015). 

\bibitem{Polin16} R. Jeanneret, D. O. Pushkin, V. Kantsler, and M. Polin, Entrainment dominates the interaction of microalgae with micron-sized objects, Nat. Commun. 7, 12518 (2016). 

\bibitem{Polin18} A. J. T. M. Mathijssen, R. Jeanneret, and M. Polin, Universal entrainment mechanism controls contact times with motile cells, Phys. Rev. Fluids 3, 033103 (2018). 

\bibitem{Yeomans17} H. Shum and J. M. Yeomans, Entrainment and scattering in microswimmer-colloid interactions, Phys. Rev. Fluids 2, 113101 (2017).

\bibitem{Ortlieb19} L. Ortlieb, S. Rafai, P. Peyla, C. Wagner, and T. John, Statistics of colloidal suspensions stirred by microswimmers,  Phys. Rev. Lett. 122, 148101 (2019).

\bibitem{Gibaud20} A. Lagarde, N. Dages, T. Nemoto, V. Demery, D. Bartolo, and T. Gibaud, Colloidal transport in bacteria suspensions: from bacteria collision to anomalous and enhanced diffusion, Soft Matter 16, 7503 (2020).

\bibitem{Leonardo11} L. Angelani, C. Maggi, M. L. Bernardini, A. Rizzo, and R. Di Leonardo, Effective interactions between colloidal particles suspended in a bath of swimming cells, Phys. Rev. Lett. 107, 138302 (2011).

\bibitem{Reichhardt14} D. Ray, C. Reichhardt, and C. J. Olson Reichhardt, Casimir effect in active matter systems, Phys. Rev. E 90, 013019 (2014).

\bibitem{Ni15} R. Ni, M.A. Cohen Stuart, and P.G. Bolhuis, Tunable long range forces mediated by self-propelled colloidal hard spheres, Phys. Rev. Lett. 114, 018302 (2015).

\bibitem{Ali-Naji17} M. Z. Yamchi and A. Naji, Effective interactions between inclusions in an active bath, J. Chem. Phys. 147, 194901 (2017).

\bibitem{Zhao21} F. Feng, T. Lei, and N. Zhao, Tunable depletion force in active and crowded environments, Phys. Rev. E 103, 022604 (2021).

\bibitem{Lowen15} F. Smallenburg and H. Löwen, Swim pressure on walls with curves and corners, Phys. Rev. E 92, 032304 (2015).

\bibitem{Cacciuto14} J. Harder, S. A. Mallorya, C. Tung, C. Valeriani, and A. Cacciuto, The role of particle shape in active depletion, J. Chem. Phys. 141, 194901 (2014).

\bibitem{Yang20} P. Liu, S. Ye, F. Ye, K. Chen, and M. Yang, Constraint dependence of active depletion forces on passive particles, Phys. Rev. Lett. 124, 158001 (2020). 

\bibitem{Kafri18} Y. Baek, A. P. Solon, X. Xu, N. Nikola, and Y. Kafri, Generic long-range interactions between passive bodies in an active liquid, Phys. Rev. Lett. 120, 058002 (2018).

\bibitem{Hagan12} S. R. McCandlish, A. Bhaskaran, and M. Hagan, Spontaneous segregation of self-propelled particles with different motilities, Soft Matter 8, 2527 (2012). 

\bibitem{Brady15} S. C. Takatori and J. F. Brady, A theory for the phase behavior of mixtures of active particles, Soft Matter 11, 7920 (2015).

\bibitem{Brady19} A. K. Omar, Y. Wu, Z. G. Wang, and J. F. Brady, Swimming to stability: Structural and dynamical control via active doping, ACS Nano 13, 560 (2019).

\bibitem{Cates15-2}  J. Stenhammar, R. Wittkowski, D. Marenduzzo, and M. E. Cates, Activity-induced phase separation and self-assembly in mixtures of active and passive particles,Phys. Rev. Lett. 114, 018301 (2015).

\bibitem{Frey16}  S. N. Weber, C. A. Weber and E. Frey, Binary mixtures of particles with different diffusivities demix, Phys. Rev. Lett. 116, 058301 (2016).

\bibitem{Gompper16} A. Wysocki, R. G. Winkler, and G. Gompper, Propagating interfaces in mixtures of active and passive Brownian particles, New J. Phys. 18, 123030 (2016).

\bibitem{Dolai18} P. Dolai, A. Simha, and S. Mishra, Phase separation in binary mixtures of active and passive particles, Soft Matter 14, 6137 (2018).

\bibitem{Adler67} J. Adler, and B. Templeton, The effect of environmental conditions on the motility of Escherichia coli. J. Gen. Microbiol. 46, 175–184 (1967).

\bibitem{Joanny20} E. Ilker and J.-F. Joanny, Phase separation and nucleation in mixtures of particles with different temperatures, Phys. Rev. Res. 2, 023200 (2020).

\bibitem{Joanny15}A. Y. Grosberg and J.-F. Joanny, Nonequilibrium statistical mechanics of mixtures of particles in contact with different thermostats, Phys. Rev. E 92, 032118 (2015).

\bibitem{Liebchen20} F. Hauke, H. Lowen, and B. Liebchen, Clustering-induced velocity-reversals of active colloids mixed with passive particles, J. Chem. Phys. 152, 014903 (2020).


\bibitem{Schilling}  R. L. Schilling and L. Partzsch, in Brownian Motion:An Introduction to Stochastic Processes, contributed by B. Böttcher (De Gruyter Textbook, 2012).


\bibitem{Schuss}  Z. Schuss, Brownian Dynamics at Boundaries and Interfaces: In Physics, Chemistry, and Biology (Springer, New York, 2015)

\bibitem{Fily12} Y. Fily and M. C. Marchetti, Athermal Phase Separation of Self-Propelled Particles with No Alignment, Phys. Rev. Lett. 108, 235702 (2012).

\bibitem{Catesprl2015} A. Tiribocchi, R. Wittkowski, D. Marenduzzo, and M. E. Cates, Active Model H: Scalar Active Matter in a Momentum-Conserving Fluid, Phys. Rev. Lett., 115, 188302 (2015).

\bibitem{csd} F. Peruani and M. Bär, A kinetic model and scaling properties of non-equilibrium clustering of self-propelled particles, New J. Phys., 15, 065009(2013).

\bibitem{jaydeb}J. P. Singh, S. Pattanayak, S. Mishra and J. Chakrabarti, Effective single component description of steady state structures of passive particles in an active bath, The Journal of Chemical Physics, 156,(2022).

\bibitem{jaydeb1} J. Dzubiella, J. Chakrabarti and H. Lowen, Tuning colloidal interactions in subcritical solvents by solvophobicity: Explicit versus implicit modeling, J. Chem. Phys. 131,044513(2009).

\bibitem{jaydeb2} J Chakrabarti, S Chakrabarti and H Lown, Short ranged attraction and long ranged repulsion between two solute particles in a subcritical liquid solvent, J. Phys. Condense Matter 18,81-87(2006).

\bibitem{Asakura54} S. Asakura and F. Oosawa, On Interaction between Two Bodies Immersed in a Solution of Macromolecules, J. Chem. Phys. 22, 1255(1954).

\bibitem{Lekkerkerker92} H.N.W. Lekkerkerker, W. C. K. Poon, P. N. Pusey, A. Stroobants, and P. B. Warren, Phase Behaviour of Colloid + Polymer Mixtures, Europhys. Lett. 20, 559 (1992).

\bibitem{Madan19} A.S. Vishen, J. Prost, M. Rao, Breakdown of effective temperature, power law interactions, and self-propulsion in a momentum-conserving active fluid, Phys. Rev. E. 100 (6), 062602 (2019).

\bibitem{Garcia15} R.C. Krafnick and A.E. Garcia, Impact of hydrodynamics on effective interactions in suspensions of active and passive matter, Phys. Rev. E 91, 022308 (2015).




\end{thebibliography}
\end{document}


\title{Supplementary information\\ Phase separation of passive particles in active liquids}
\author{Pragya Kushwaha$^{1}$, Vivek Semwal$^2$, Sayan Maity$^1$, Shradha Mishra$^{2}$, Vijayakumar Chikkadi$^{1}$}

\affiliation{
$^{1}$ Indian Institute of Science Education and Research Pune, India 411008\\
$^{2}$ Indian Institute of Technology (BHU) Varanasi, India 221005\\
}

\maketitle

\section{Methods}
\subsection{Experiments}
The active suspensions were prepared using E.coli cells (U5/41 type strain). The bacteria were grown overnight at $37^{\circ}C$ in LB agar plate containing $1\%$ tryptone, $1\%$ NaCl, $0.5\%$ yeast extract and $1.5\%$ agar. A single colony of E. coli was added to $10~ml$ of LB broth and kept at $37^{\circ}C$  until OD$_{600}$ (optical density at $600~nm$ wavelength) reached a value of $1.3$. Bacterial cells were then harvested and washed three times with motility media  ($10~mM$ potassium phosphate ($pH~7.0$), $0.1~mM$ EDTA , $0.002\%$ Tween-20 and $50~mM$ L-Serine) by centrifugation at $3000$ rpm for $5$ minutes at room temperature to remove the traces of LB broth. The pellet was later resuspended in motility media to get desired concentrations. The observation chamber was created using a circular cavity of size $1~cm$ and $100~\mu m$ deep, which was glued to a PEG coated coverslip using double-sided tape. The bacteria density of the sample with OD$_{600}=1$ was estimated to contain $6\times10^9$ cells/ml $(b_0)$. The density of bacteria in our experiments was fixed at 10$b_0$, which was well below the density threshold for the onset of collective motion.  

\subsection{Simulations}

The simulations were performed using a binary mixture with $N_a$ small active particles of radius $a_a$ and $N_p$ big passive particles of radius $a_p$ ($a_p > a_a$) moving on a two dimensional frictional substrate. The active particles are associated with a self propulsion speed ${v}$ and an orientation unit vector $\hat{v}_i = (cos\theta_i; sin\theta_i)$, where $\theta_i$ is the angle between the velocity vector and a reference direction. The motion of active Brownian particles (ABP) is governed by the following Langevin equations:
\begin{equation}	 \tag{S1}
\frac{d{\bf r}_i}{dt}=v\hat{v}_i-\mu_1\sum_{j\neq i}{\bf F}_{ij} + \sqrt{2 D_T} {\bf \eta}_{Ti}
	\label{eq1}
\end{equation}
\begin{equation} \tag{S2}
\frac{d\theta_i}{dt}=\Gamma\sum_{i}sin(\theta_i-\theta_{ij})+\sqrt{2D_r}\eta_{ri}.
\label{eq2}
\end{equation}
Here $\mu_1$ is the mobility and ${\bf F}_{ij}$ is the force acting on particle $i$ due to particle $j$. The noise term is defined as $<\eta_{r,Ti}(t)\eta_{r,Tj}(t^{\prime})> =2D_{r,T}\delta_{ij}\delta{(t-t^{\prime})}$, $D_T$ and $D_r$ are the translational and rotational diffusion constants of active particles and $\Gamma$ is the magnitude of alignment and $\theta_{ij}=\arg{({\bf r}_i - {\bf r}_j )}$. The persistence length $l_p = v/D_r$ of active particles is constant in our simulations, it is set at $l_p=20a_a$. The other constants in our simulations are $D_T=0.005$ and $\Gamma=1$ \\

The equation of motion for passive particles is
\begin{equation} \tag{S3}
\frac{d{\bf r}_i}{dt}=\mu_2\sum_{j\neq {i}}\mathbf{F}_{ij},
	\label{eq3}
\end{equation}
where $\mu_2$ is the mobility of passive particles. There is no translational noise in Eq.~3, so the dynamics of passive particles is only due to interaction force. We choose the mobility of both species to be the same i.e., $\mu_1 = \mu_2$. Particles interact through short ranged soft repulsive forces $\mathbf{F}_{ij} = F_{ij} \hat{r}_{ij}$, where $F_{ij} = k(a_i +a_j-r_{ij})$ when $r_{ij}\leq (a_i+a_j)$ and $F_{ij} = 0$ otherwise; $r_{ij}=|r_i-r_j|$ and $k$ is a constant. The elastic time scale in our system is defined by $\left(\mu k\right)^{-1} = (150)^{-1}$.\\

\section{Supplementary figures}
\subsection{Peclet number estimation}
\begin{figure}[h]
\begin{tabular}{llll}
\includegraphics[height=0.3\textwidth]{ 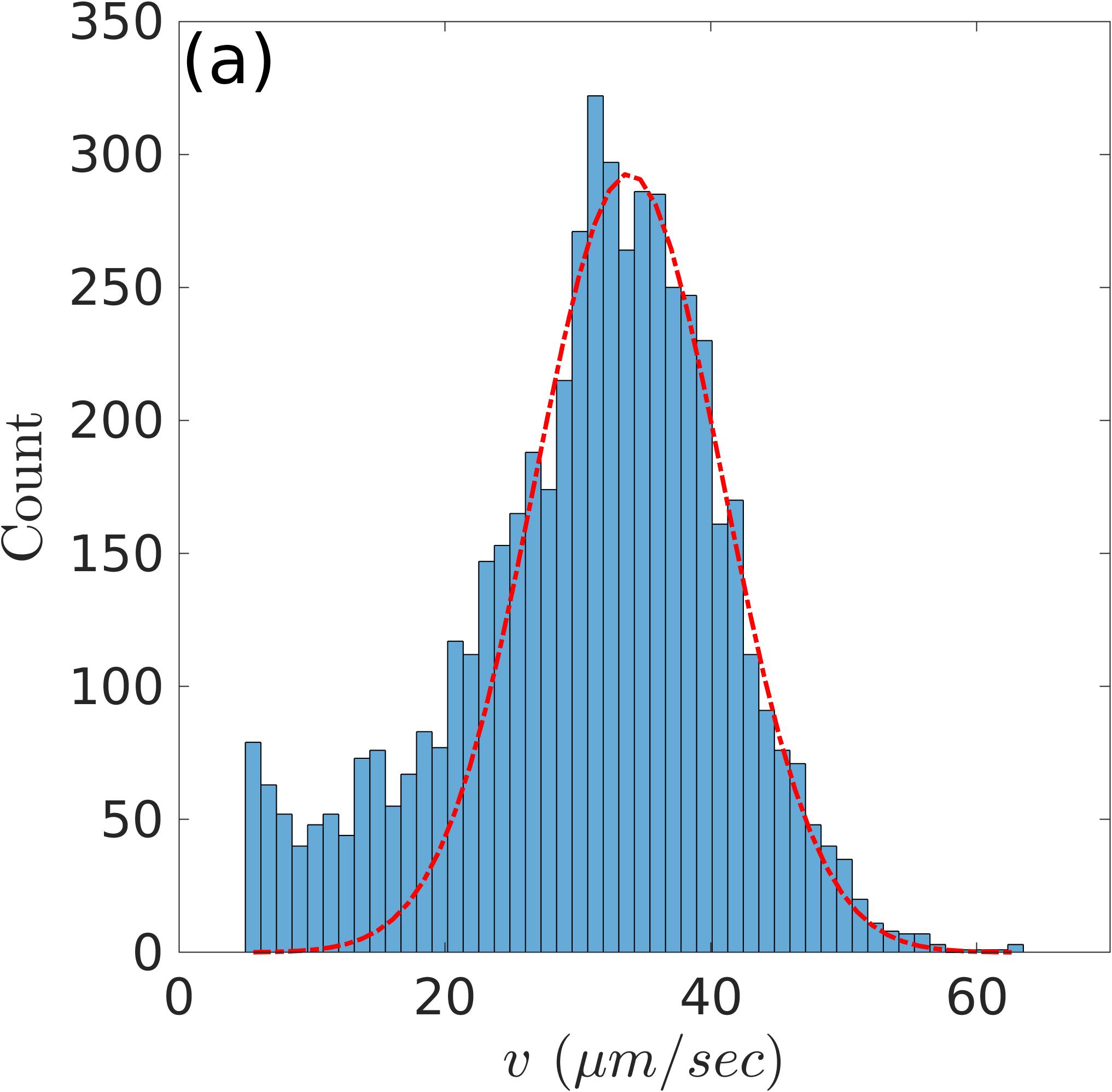}&
\includegraphics[height=0.3\textwidth]{ 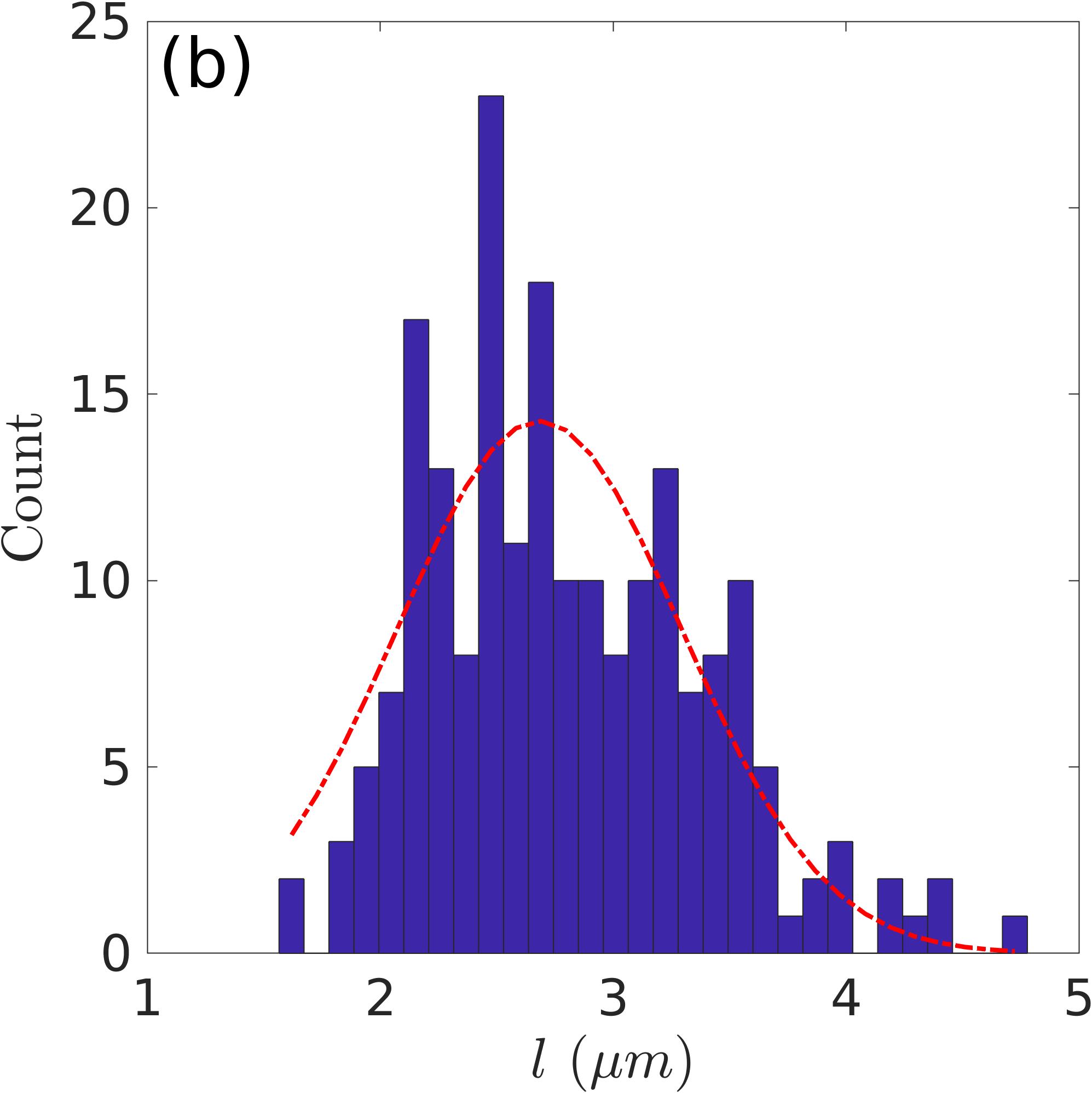}&
\includegraphics[height=0.3\textwidth]{ 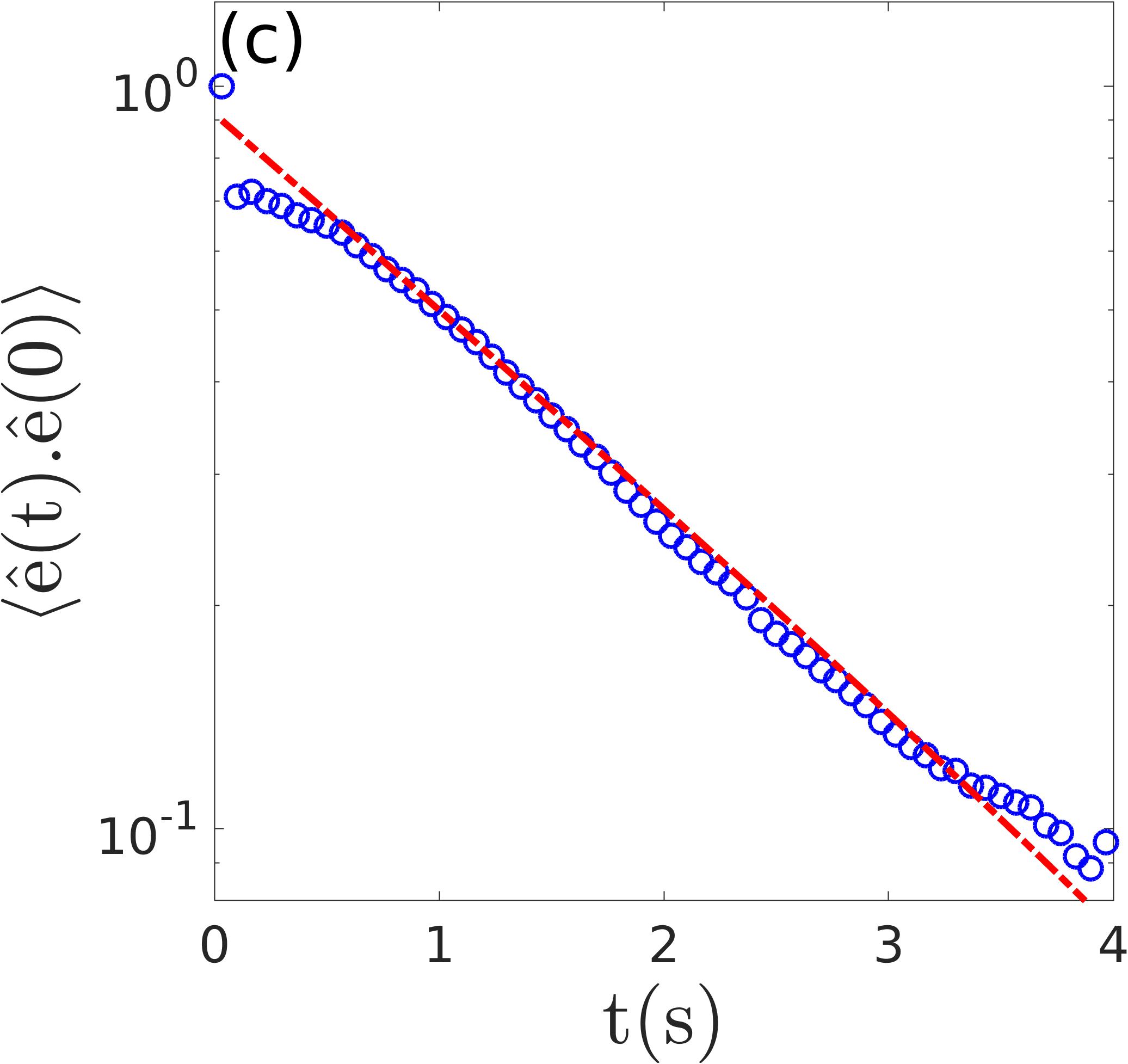}&
\end{tabular}
\caption{ (a) Histogram of bacteria velocities. The average velocity is $<v>=33.84\pm 9.98~ \mu m/s$. (b) Histogram of the size of the bacteria. The size of the bacteria is its length along the longer axis. The average length of the cells is $<l>=2.68\pm0.86~\mu m $. (c) The rotational diffusion time of the bacteria was estimated from their normalised velocity auto-correlation function. The dashed line is an exponential fit to the data, which gives a rotational diffusion time of $\tau_r\sim 1.67~s$.}
\label{figs1}
\end{figure}





\section{Supplementary figures}

\subsection{Steady state in experiments}
\begin{figure}[h]
\centering
\includegraphics[height=0.3\textwidth]{ 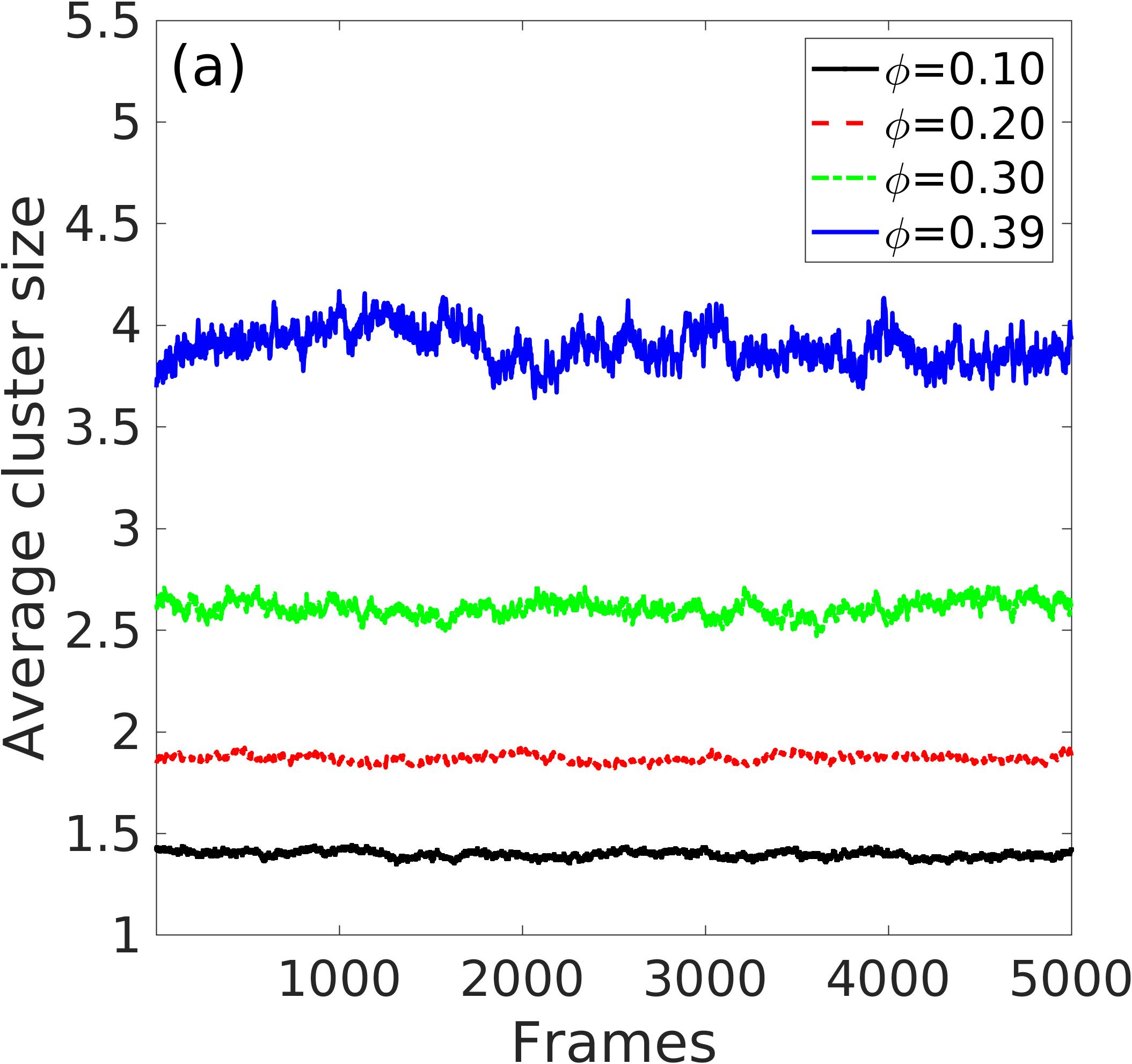}
\includegraphics[height=0.3\textwidth]{ 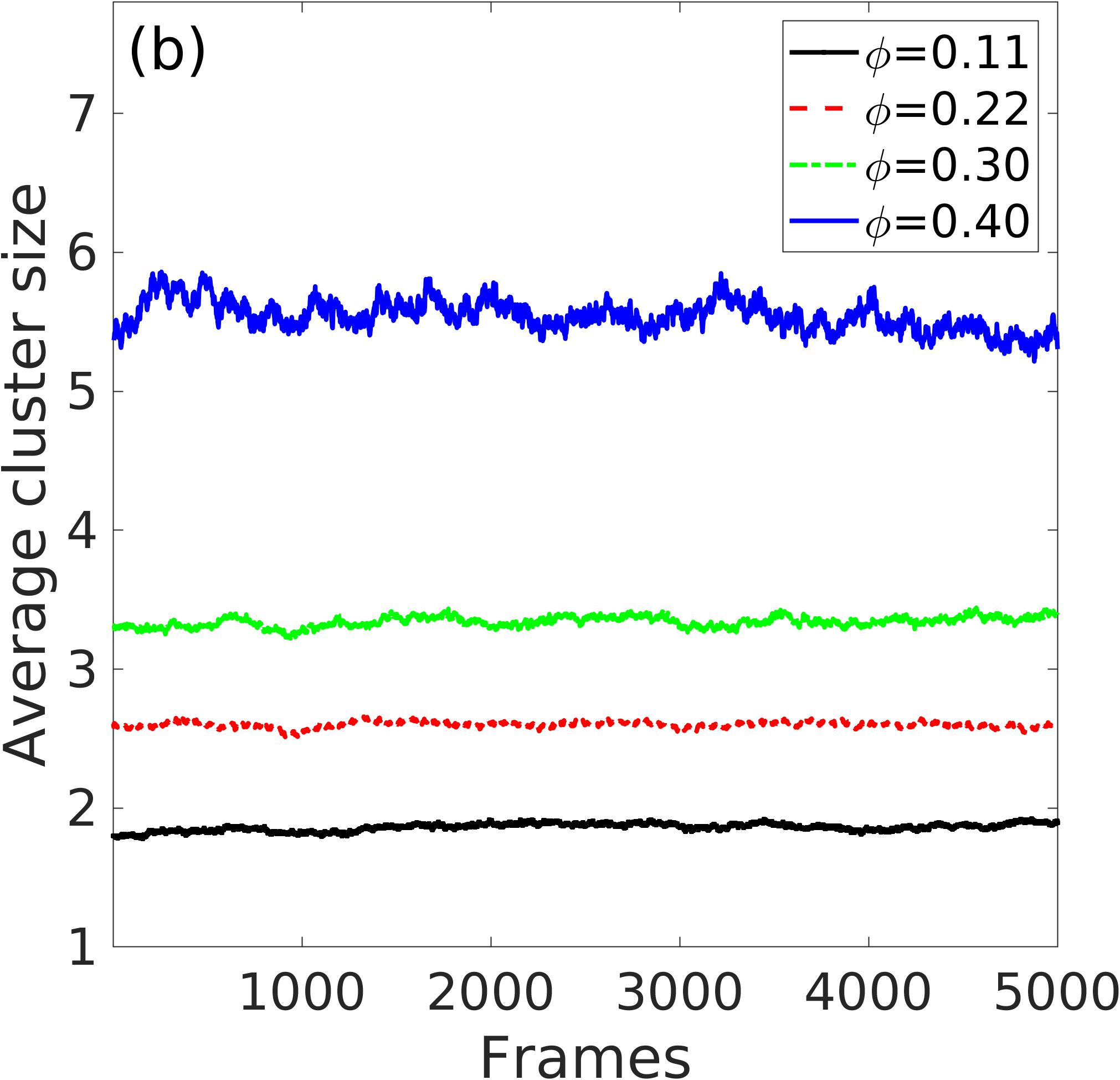}
\includegraphics[height=0.3\textwidth]{ 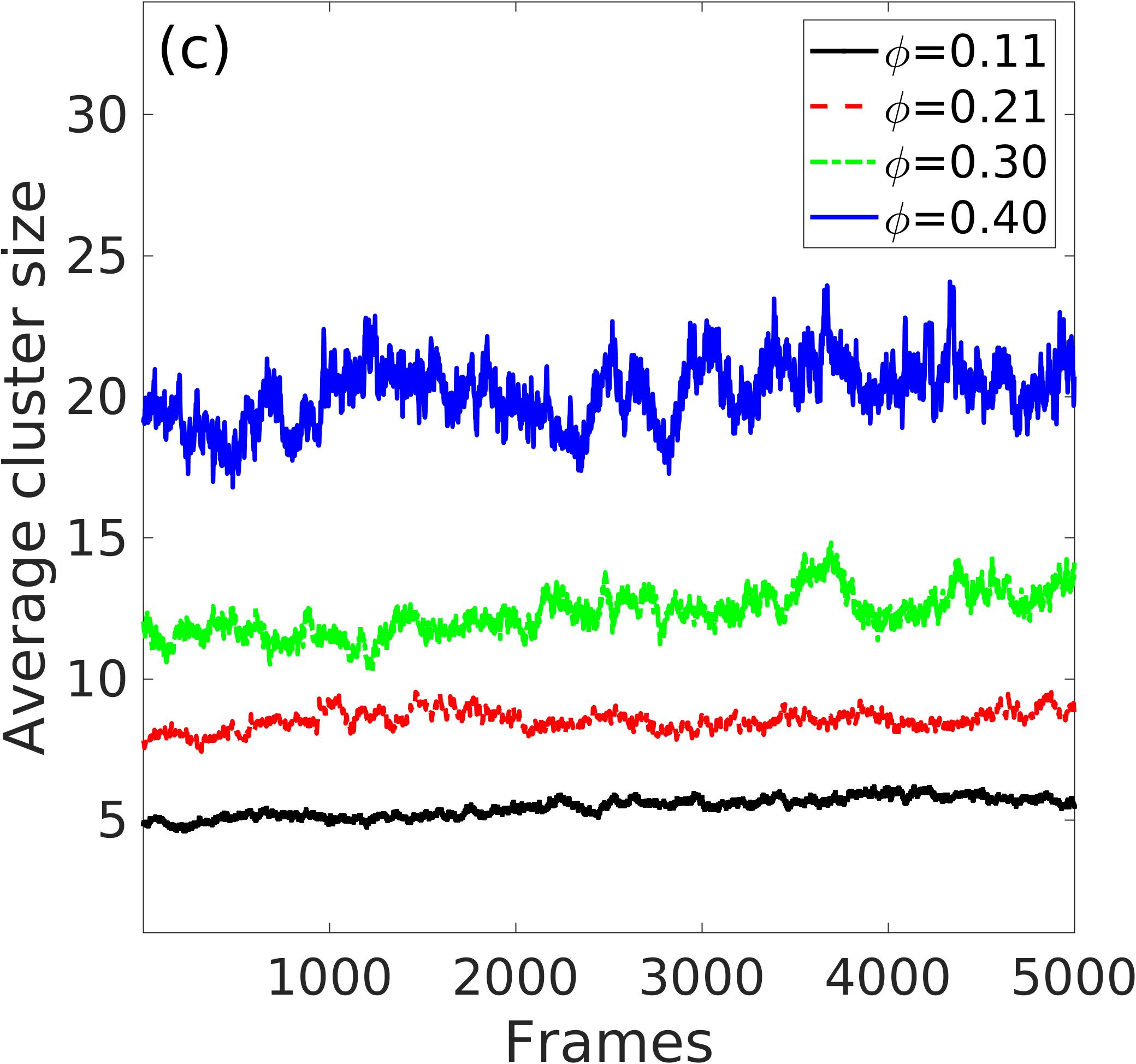}
\caption{ The average cluster size in our system as a function of frame number for a range of area fractions from $\phi\sim0.1-0.4$ and at $S\sim 2.5$ (a), $S \sim 3.5$ (b) and $S\sim 5.5$ (c). The steady state of the system is evident from these figures. The total duration of the measurement was $500~s$}
\label{figs6}
\end{figure}
\newpage
\subsection{Calculation of effective potential}

\begin{figure}[h]
    \centering
    \includegraphics[height=0.15\textwidth]{ 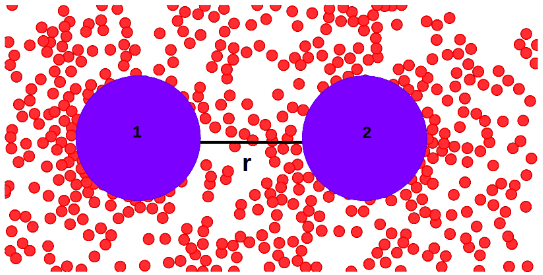}
    \caption{Snapshot of the part of the system to calculate the potential. The two bigger particles are passive particles with the left one marked as particle $1$ and the right one marked $2$ with positions ${\bf r}_1$ and ${\bf r}_2$, respectively. The red circles are ABPs. The line shows the surface to surface distance $r$ between two passive particles }
    \label{numberfluctuation}
\end{figure}

\subsection{Density of active particles in the vicinity of an isolated passive particle in simulations}

\begin{figure}[h]
\centering
\includegraphics[height=0.23\textwidth]{ 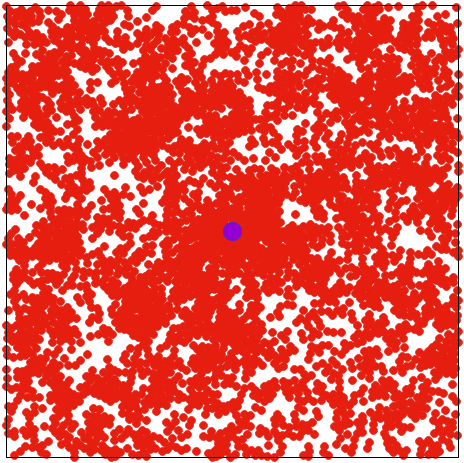}
\includegraphics[height=0.23\textwidth]{ 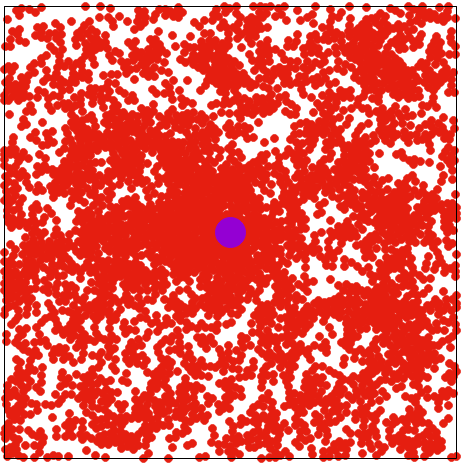}
\includegraphics[height=0.23\textwidth]{ 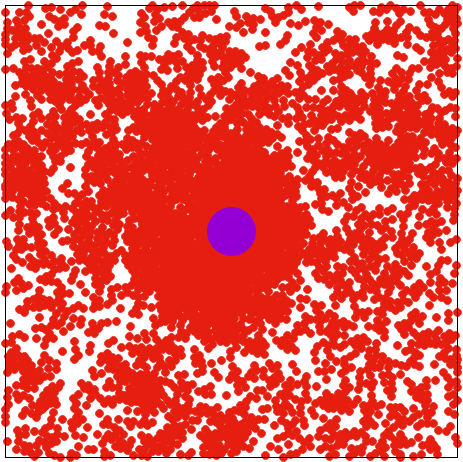}
\includegraphics[height=0.23\textwidth]{ 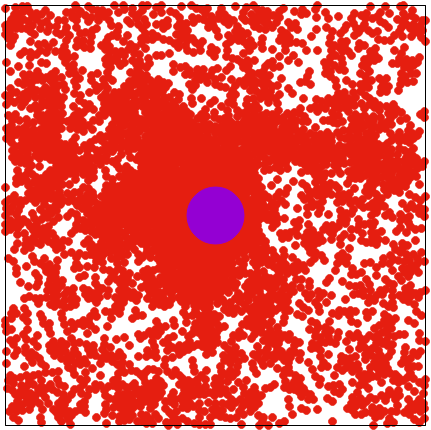}
\caption{ Density fluctuations of active particles due to a single passive particle in the center of the system. Instantaneous snapshots of the system for four size ratios $S=3$, $5$, $8$ and $10$, from left to right. Many such configurations are used to calculate the density correlations $C(r)$. Red particles are ABP's and blue particle at the center is the bigger passive particle.}
\label{snapshotcenter}
\end{figure}
\subsection{Number fluctuations of active particles in the vicinity of an isolated passive particle in simulations}

\begin{figure} [h]
\centering
\includegraphics[height=0.3\textwidth]{ 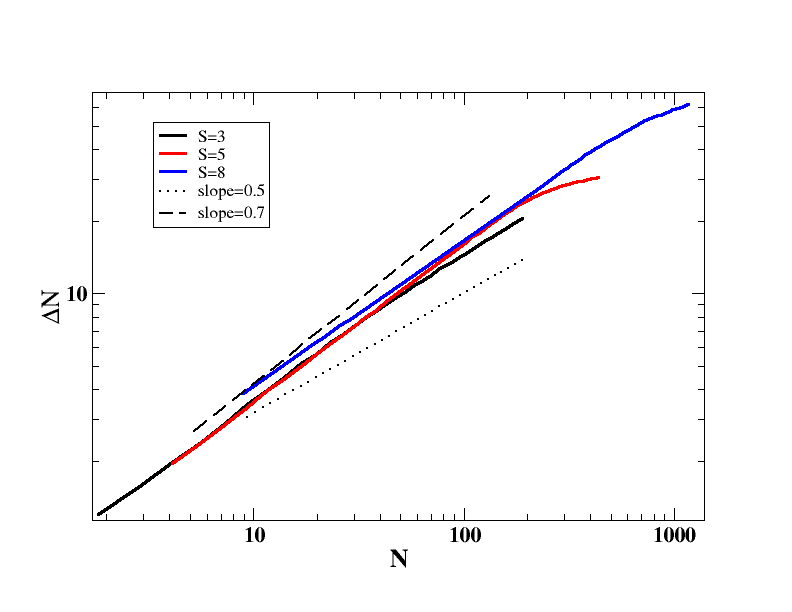}
\caption{Number fluctuation $\Delta N$ {\em vs.} $N$ for different size ratios. The slopes of dotted line is 0.5 and dashed line is 0.7}
\label{numberfluctuation}
\end{figure}

For calculating number fluctuation, we start with an annular disc with an inner radius the radius of the passive particle and outer radius is varied. The mean and variance of number of ABP's are calculated for different outer radius of the disc. The same is repeated for three different sizes of the passive particle or for three different size ratios $S=3$, $5$ and $8$. In the Fig.~\ref{numberfluctuation} we show the plot of $\Delta N$ {\em vs.} $N$ for three sizes $S=3$, $5$ and $8$. For all the cases the graph is power law with $\Delta N \simeq N^{\alpha}$, where $\alpha \simeq 0.7$ for moderate $N$ for all $S$ and starts to deviate for large $N$. The deviation appears at relatively larger $N$ on increasing size ratio.  Hence increasing the size of passive particle increases the stretch of density fluctuation of ABP's and it might be one of the factor to introduce a long ranged attraction among passive particles for larger size ratio. 

\subsection{Correlations of density fluctuations of bacteria in the vicinity of an isolated colloidal particle in experiments}

\begin{figure} [h]
\centering
\includegraphics[height=0.35\textwidth]{ 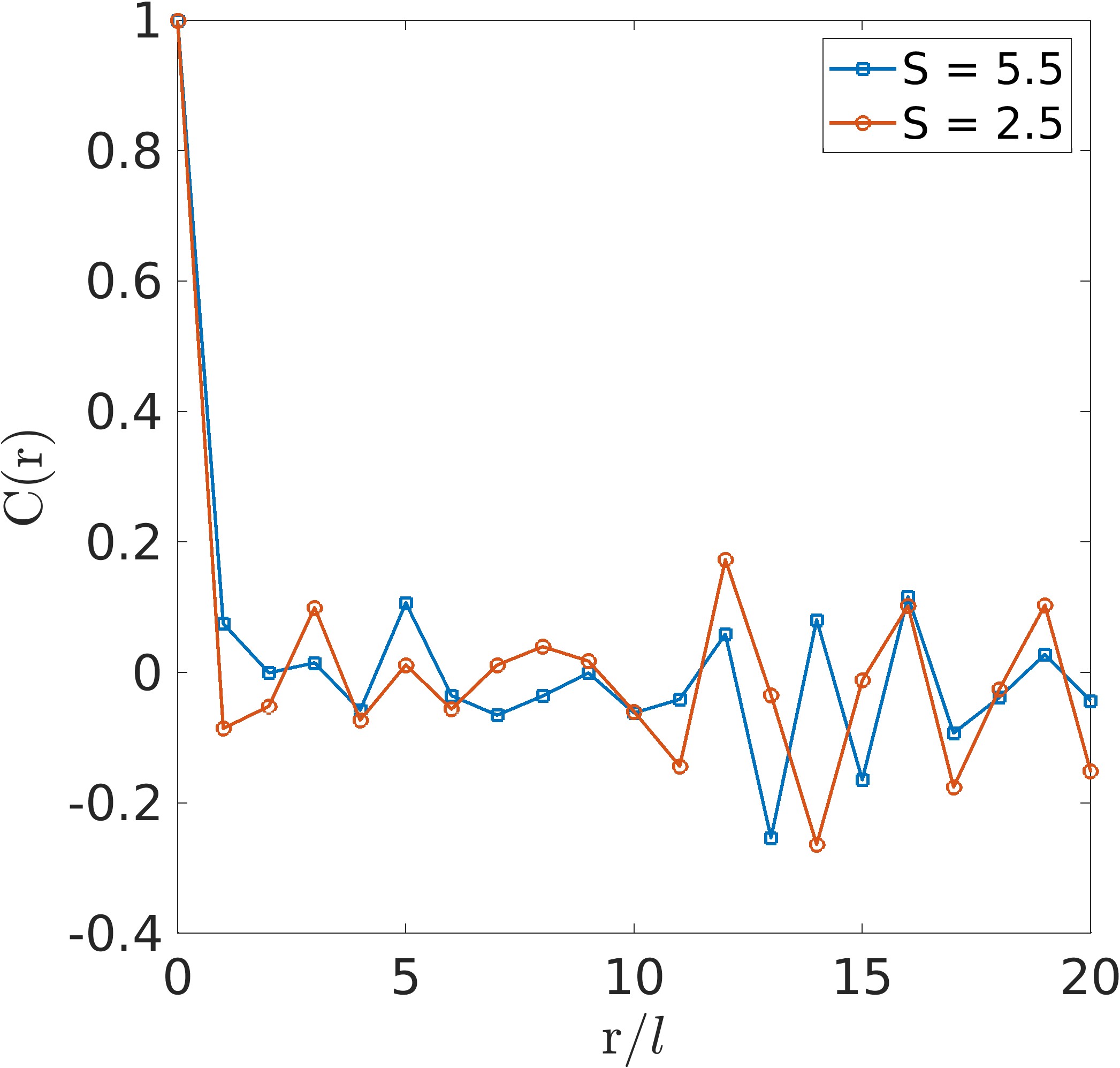}
\caption{The correlations of density fluctuations $C(r)$, as defined in the main text, is shown for two different size ratios $S\sim2.5$ and $5.5$. The x-axis is scaled by the size of the bacteria $l$. It is clear from these results that the density fluctuations are suppressed in experiments.}
\label{numberfluctuation}
\end{figure}